\documentclass[12pt]{iopart}
\usepackage{braket}
\usepackage[dvips]{graphicx}
\usepackage[spanish,english]{babel}
\usepackage{iopams}
\begin{document}

\title[Damped Driven Coupled Oscillators]
{Damped driven coupled oscillators:
entanglement, decoherence and the classical limit}

\author{R.~D.~Guerrero Mancilla, R.~R.~Rey-Gonz\'alez and 
K.~M.~Fonseca-Romero}

\address{\centerline{\emph{Grupo de \'Optica e Informaci\'on Cu\'antica,
Departamento de F\'{\i}sica,}} \centerline{\emph{Universidad
Nacional de Colombia - Bogot\'a }}} \ead{rdguerrerom@unal.edu.co}
\ead{rrreyg@unal.edu.co} \ead{kmfonsecar@unal.edu.co}

\begin{abstract}
\noindent{\footnotesize 
The interaction of (two-level) Rydberg atoms with dissipative 
QED cavity fields can be described classically or 
quantum mechanically, even for very low temperatures 
and mean number of photons, provided the damping constant is large enough. 
We investigate the quantum-classical border, the entanglement
and decoherence of an analytically solvable model,  
analog to the atom-cavity system, in which the
atom (field) is represented by a (driven and damped) harmonic oscillator.
The maximum value of entanglement is shown to depend on the initial state
and the dissipation-rate to coupling-constant ratio. 
While in the original model the atomic entropy never
grows appreciably (for large dissipation rates), in our model it reaches 
a maximum before decreasing. Although both models predict small values of
entanglement and dissipation, for fixed times of the order of the inverse 
of the coupling constant and large dissipation rates, these quantities
decrease faster, as a function of the ratio of the 
dissipation rate to the coupling constant, in our model.}

{\thanks{} This research was partially funded by DIB and Facultad de 
Ciencias, Universidad Nacional de Colombia.}
\end{abstract}


\maketitle

\section{Introduction}

One expects quantum theory to approach to the classical theory, 
for example in the singular limit of a vanishing Planck's 
constant, $\hbar\rightarrow 0$, or for large quantum 
numbers. However, dissipative systems can bring forth some surprises:
for example, QED (quantum electrodynamics) cavity fields interacting
with two-level systems, may exhibit classical or quantum behavior,
even if the system is kept at very low temperatures and if the mean 
number of photons in the cavity is of the order of 
one \cite{Kim1999a,Raimond2001a}, depending on the strength of the 
damping constant. Classical behavior, in this context refers to the 
unitary evolution of one of the subsystems, as if the other subsystem 
could be replaced by a classical driving.
In QED cavities, the atom, which enters in one of the
relevant Rydberg states (almost in resonance with the field
sustained in the cavity), conserves its purity and suffers a unitary 
rotation inside the cavity -- exactly as if it were controlled by a 
classical driving field -- without entangling with the electromagnetic 
field. This unexpected behavior was analyzed in reference 
\cite{Kim1999a} employing several short-time approximations, and it 
was found that in the time needed to rotate the atom, its state remains
almost pure.

Other driven damped systems, composed by two (or more) subsystems
can be readily identified. Indeed, in the last years there has 
been a fast development of quite different physical systems
and interfaces between them, including electrodynamical cavities
\cite{cavities1,cavities8}, superconducting circuits
\cite{SCCircuits1,SCCircuits4}, confined electrons
\cite{CElectrons4,CElectrons5,CElectrons6} and nanoresonators
\cite{NMResonators1,NMResonators5,NMResonators6}, on which it is
possible to explore genuine quantum effects at the level of a few
excitations and/or in individual systems. For
instance, the interaction atom-electromagnetic field is exploited in
experiments with trapped ions \cite{TrappedIons2,TrappedIons3},
cavity electrodynamics and ensembles of atoms interacting with
coherent states of light \cite{Polzik}, radiation pressure over
reflective materials in experiments coupling the mechanical motion
of nanoresonators to light \cite{NMResonators6}, and the coupling of
cavities with different quality factors in the manufacturing of 
more reliable Ramsey zones \cite{Haroche07}. In many of
these interfaces it is possible to identify a system which couples
strongly to the environment and another which couples weakly. For 
example, in the
experiments of S. Haroche the electromagnetic field decays
significantly faster \cite{Haroche96} (or significantly slower
\cite{Haroche07}) than the atoms, the quality factor $Q$ of the
nanoresonators is much smaller than that of the cavity, and the newest
Ramsey zones comprise two coupled cavities of quite different $Q$.  
Several of these systems therefore, can be modelled as coupled
harmonic oscillators, one which can be considered dissipationless.

In this contribution we study an exactly solvable system, composed
of two oscillators, which permits the analysis of large times,
shedding additional light on the classical-quantum border. 
Entanglement and entropy, as measured by concurrence and linear 
entropy, are used to tell ``classical'' from quantum effects.

\section{The model}
The system that we consider in this manuscript comprises two
oscillators of natural frequencies $\omega_1$ and $\omega_2$,
coupled through an interaction which conserves the (total) number of
excitations and whose coupling constant abruptly changes from zero
to $g$ at some initial time, and back to zero at some final time.
We take into account that the second
oscillator loses excitations at the rate $\gamma$, through a
phenomenological Liouvillian of Lindblad form, corresponding to zero
temperature, in the dynamical equation of motion \cite{Louisell1973a}.
Lindblad superoperators are convenient because they preserve important
characteristics of physically realizable states, namely hermiticity,
conservation of the trace and semi-positivity \cite{Lindblad1976a}.
In order to guarantee the presence of excitations, the second 
oscillator is driven by a classical resonant field.

The interaction can be considered to be turned on (off) in the remote 
past (remote future) if it is always present (coupled Ramsey zones or
nanoresonators coupled to cavity fields), or can really be present for
a finite time interval (for example in atoms travelling through 
cavities). The initial states of the coupled oscillators also depend on 
the experimental setup, varying from the base state of the compound 
system to a product of the steady state of the coupled damped oscillator 
with the state of the other oscillator. Since we want to make comparisons 
with Ramsey zones, the choices in the formulation of this model have been 
inspired by the analogy with the atom-cavity system, for a cavity --kept 
at temperatures of less than 1K-- whose lifetime is much shorter than 
the lifetime of Rydberg states, allowing us to ignore the Lindblad 
operator characterizing the atomic decay process. The first oscillator 
therefore is a cartoon of the atom, at least in the limit where only its 
first two states are significantly occupied, while the second oscillator 
corresponds to the field.

All the ingredients detailed before can be summarily put into the
Liouville-von Neumann equation for the density matrix $\hat{\rho}$
of the total system
\begin{equation}
\label{eq:lvneq} \fl \frac{d{\hat \rho}}{dt}=
-\frac{i}{\hbar}[{\hat H},{\hat \rho}]
+ \gamma
(2{\hat a}{\hat \rho}{\hat a}^{\dagger}
-{\hat a}^{\dagger}{\hat a}{\hat \rho}
-{\hat \rho}{\hat a}^{\dagger}{\hat a})
\end{equation}
where $\hat{H}$ is the total Hamiltonian of the system and the
second term of the rhs of (\ref{eq:lvneq}) is the Lindblad
superoperator which accounts for the loss of excitations
of the second oscillator.
In absence of the coupling with the first oscillator, the inverse
of twice the dissipation rate $\gamma$ gives the mean
lifetime of the second oscillator. The first
two terms of the total Hamiltonian
\begin{equation}
 \fl
\hat{H} = \hbar\omega_{1}{\hat b}^{\dagger}{\hat b}
+\hbar\omega_{2}{\hat a}^{\dagger}{\hat a}
+ \hbar g \left(\Theta(t)-\Theta(t+T)\right)
({\hat a}^{\dagger}{\hat b}+{\hat a}{\hat b}^{\dagger})
+ i \hbar\epsilon
(e^{-i\omega_D t}{\hat a}^{\dagger}-e^{i\omega_D t}{\hat a}),
\end{equation}
are the free Hamiltonians of the two harmonic oscillators; the next
term, which is modulated by the step function $\Theta(t)$, is the
interaction between them and the last is the driving.
The bosonic operators of creation $\hat b$ ($\hat a$) and
annihilation $\hat b^\dag$ ($\hat a^\dag$) of one excitation of the
first (second) oscillator, satisfy the usual commutation relations.
From here on we focus on the case of resonance,
$\omega_1=\omega_2=\omega_D=\omega$. The interaction time $T$ is left
undefinite until the end of the manuscript, where we compare our
results with those of the atom-cavity system.

\section{Dynamical evolution}
The solution of the dynamical equation (\ref{eq:lvneq}) can be written
as
\begin{equation}
\label{ansatz} \fl
 \hat{\rho}(t) = {\mathcal D}(\beta(t),\alpha(t))
\tilde{\rho}(t) {\mathcal D}^\dag(\beta(t),\alpha(t)),
\end{equation}
where ${\mathcal D}(\beta(t),\alpha(t))$ is the two-mode displacement
operator,
\begin{equation*} \fl
 {\mathcal D}(\beta(t),\alpha(t)) =
 {\mathcal D}_1(\beta(t))  {\mathcal D}_2(\alpha(t))
= e^{\beta(t) \hat{b}^\dag - \beta^*(t) \hat{b}}
e^{\alpha(t) \hat{a}^\dag - \alpha^*(t) \hat{a}},
\end{equation*}
and $\tilde{\rho}(t)$ is the total density operator in the interaction
picture defined by equation (\ref{ansatz}).
By replacing (\ref{ansatz}) into (\ref{eq:lvneq}), and employing the
operator identities
\begin{equation*} \fl
\frac{d}{dt}{\mathcal D}(\alpha) = \left( -\frac{\alpha^*
\dot{\alpha}-\dot{\alpha}^* \alpha}{2} + \dot{\alpha} \hat{a}^\dag -
\dot{\alpha}^* \hat{a}\right) {\mathcal D}(\alpha) = {\mathcal
D}(\alpha) \left( \frac{\alpha^* \dot{\alpha}-\dot{\alpha}^*
\alpha}{2} + \dot{\alpha} \hat{a}^\dag - \dot{\alpha}^*
\hat{a}\right),
\end{equation*}
with the dot designating the time derivative as usual,  we are able to
decouple the dynamics of the displacement operators, obtaining the
following dynamical equations for the labels $\alpha$ and $\beta$
\begin{eqnarray}
\label{eq:dDisplacementdt}  \fl \frac{d}{dt} \left(
\begin{array}{c}
\alpha \\ \beta
\end{array}
\right ) =
\left(
\begin{array}{rc}
-\gamma-i\omega & -i g \\
-i g  & -i\omega
\end{array}
\right ) \left(
\begin{array}{c}
\alpha \\ \beta
\end{array}
\right)
+\left(
\begin{array}{c}
\epsilon e^{-i\omega t} \\ 0
\end{array}
\right),
\end{eqnarray}
for times between zero and $T$.
On the other hand, the Ansatz (\ref{ansatz}) also provides the
equation of motion for $\tilde{\rho}(t)$, which turns out to be very
similar to (\ref{eq:lvneq}) but with the hamiltonian $\tilde{H}=
\hat{H}(\epsilon=0)$, that is, without driving. The separation
provided by our Ansatz is also appealing from the point of view
of its possible physical interpretation, because the effect of the
driving has been singled out, and quantum (entangling and purity)
effects are extracted from the displaced density operator 
$\tilde{\rho}(t)$.

The two oscillators interact after the second oscillator reaches its
stationary coherent state
\begin{equation}
\label{ec:steadystate1}  \fl\hat{\rho}_2(t) = \textrm{tr}_1
\hat{\rho}(t) = \Ket{\frac{\epsilon}{\gamma} e^{-i\omega t}}
\Bra{\frac{\epsilon}{\gamma} e^{-i\omega t}},
\end{equation}
as can be verified by solving (\ref{eq:dDisplacementdt}) with the
interaction turned off. If we want a mean number of excitations of
the order of one then the driving amplitude must satisfy
$\epsilon\approx \gamma$, and thereby the larger the dissipation is,
the larger the driving is to be chosen. At zero time, when the
oscillators begin to interact, the state of the total system is
separable with the second oscillator state given by
(\ref{ec:steadystate1}). The first oscillator, on the other hand,
begins in a pure state which we choose as a linear combination of
its ground and first excited states (again inspired on the analogy
with the atom-cavity system). Thus, the initial state
$\hat{\rho}(0)$ given by
\begin{equation} \fl
\label{eq:InitialState}
{\mathcal D}\left(0,\frac{\epsilon}{\gamma}\right) \underbrace{
\left(\cos(\theta)\ket{0}+\sin(\theta)\ket{1}\right)
\left(\cos(\theta)\bra{0}+\sin(\theta)\bra{1}\right) \otimes
\ket{0}\bra{0}}_{\tilde{\rho}(0)} {\mathcal
D}^\dag\left(0,\frac{\epsilon}{\gamma}\right),
\end{equation}
corresponds to a state of the form described by equation (\ref{ansatz})
with $\beta(0)=0$ and $\alpha(0) ={\epsilon}/{\gamma}$. At later
times, the solution maintains the same structure, but --as can be
seen from the solution of (\ref{eq:dDisplacementdt}) -- the labels of
the displacement operators evolve as follows
\begin{eqnarray}
\fl \label{eq:alpha}
\alpha(t) =
\epsilon e^{-\frac{1}{2}(\gamma + 2 i\omega )t}
\left\{ \frac{1}{\gamma }  \cos\left(\tilde{g} t\right) +
 \frac{\sin\left(\tilde{g} t\right)}{2\tilde{g}}
\right\},
\\
\label{eq:beta}\fl
\beta(t)  = -\frac{i e^{-i\omega t} \epsilon }{g}
+ \frac{i \epsilon}{g} e^{-\frac{1}{2}(\gamma + 2 i \omega )t}
\left\{\cos \left(\tilde{g} t\right)
+\left(-2 g^2 +\gamma^2\right)
\frac{\sin\left(\tilde{g}t\right)}{2\gamma\tilde{g}}\right\},
\end{eqnarray}
where we have defined the new constant $\tilde{g}  =\frac{1}{2}
\sqrt{4g^2-\gamma^2 }.$ We
employ $\tilde{g}$, which also appears in the solution of the 
displaced density operator, to define three different regimes: 
underdamped ($\tilde{g}^2> 0$), 
critically damped ($\tilde{g}^2 =0$) and 
overdamped ($\tilde{g}^2 <0$) regime. It is important to notice
that there is no direct connection with the quality factor of 
the damped oscillator: it is possible to have physical systems 
in the overdamped regime defined here even with relatively large
quality factors, if the interaction constant $g$ is much smaller
than $\omega$, the frequency of the oscillators.

The inspection of the equations
(\ref{eq:alpha}) and (\ref{eq:beta}), allows one to clearly identify
the time scale $2/\gamma$, after which the stationary state is
reached and the state of the first oscillator just rotates with
frequency $\omega$ and have a mean number of excitations equal to
$\epsilon^2/g^2$. The doubling of the damping time of the second
oscillator, from $1/\gamma$ in the absence of interaction to
$2/\gamma$, in the underdamped regime, can be seen as an instance of
the shelving effect \cite{Wineland1975a}. The first oscillator,
which in absence of interaction, suffers no damping, it is now
driven and damped. It can be thought that the excitations remain
half of the time on each oscillator, and that they decay with a
damping constant $\gamma$, thereby leading to an effective damping
constant of $\gamma/2$. An interesting feature of the solution is
that the displacement of the second oscillator goes to zero, in the
stationary state. In the stationary state, the first oscillator 
evolves as if it were driven
by a classical field $-i\hbar\epsilon \exp(-i\omega t)$ and damped with a
damping rate $g$, without any interaction with a second oscillator. 
More generally speaking, we remark that from the
point of view of the first oscillator, the evolution of its
displacement operator happens as if there were damping but no
coupling, and the driving were of the form $\hbar g (\beta-i\alpha)$,
or, in terms of the parameters of the problem,
\begin{equation}
\label{eq:effectivedrive}
\fl
F(t)  = -i\hbar\epsilon e^{-i\omega t}
-i\hbar\epsilon e^{-(\gamma/2+i\omega) t}
\left(
\left(\frac{g}{\gamma}-1\right)\cos(\tilde{g}t)
+\frac{2g^2+g\gamma-\gamma^2}{2\gamma\tilde{g}}\sin(\tilde{g}t)
\right).
\end{equation}
This behavior is particularly relevant in the following extreme
case, whose complete solution depends only on the displacement
operators. If the initial state of the first oscillator is the
ground state then $\tilde{\rho}$ does not evolve in time, i.~e. it
remains in the state $\ket{00}$, and the total pure and separable
joint state is
\begin{equation} \fl
\rho(t) =
\ket{\beta(t)}\bra{\beta(t)}\otimes\ket{\alpha(t)}\bra{\alpha(t)}.
\end{equation}

Even in the more general case considered here, corresponding to the 
initial state (\ref{eq:InitialState}), the solution of $\tilde{\rho}(t)$ 
possesses only a few non-zero elements.
If we write the total density operator as
\begin{equation}
\tilde{\rho}(t) = \sum_{i_1 i_2 j_1 j_2} \tilde{\rho}_{i_1 i_2}^{j_1 j_2}
\ket{i_1 i_2} \bra{j_1 j_2},
\end{equation}
we can arrange the elements corresponding to zero and one excitations in 
each oscillator, as the two-qubit density matrix
\begin{equation}
\left(
\begin{array}{llll}
 \tilde{\rho}_{00}^{00}(t) & \tilde{\rho}_{00}^{01}(t) 
                           & \tilde{\rho}_{00}^{10}(t) & 0 \\
 \tilde{\rho}_{01}^{00}(t) & \tilde{\rho}_{01}^{01}(t) 
			   & \tilde{\rho}_{01}^{10}(t) & 0 \\
 \tilde{\rho}_{10}^{00}(t) & \tilde{\rho}_{10}^{01}(t) 
			   & \tilde{\rho}_{10}^{10}(t) & 0 \\
 0 & 0 & 0 & 0
\end{array}
\right).
\end{equation}
If we measure time in units of $g$ by defining
$t'=g t$  we have only two free parameters $\Gamma=\frac{\gamma}{g}$
and $\Omega=\frac{\omega}{g}$. The nonvanishing elements of the density
matrix, written in the underdamped case ($|\Gamma|<2$), are given
by (hermiticity of the density operator yields the 
missing non-zero elements)
\begin{eqnarray*}
\fl {  \tilde{\rho}}^{00}_{00}(t')= & 1 -\sin^2\theta e^{-\Gamma t'}
\left( \frac{4-\Gamma^2 \cos(\sqrt{4-\Gamma^2}t')}{4-\Gamma^2} -
\frac{\Gamma \sin(\sqrt{4-\Gamma^2}t')}{\sqrt{4-\Gamma^2}} \right)
\\ \fl
{ \tilde{\rho}}^{01}_{01}(t')  =&
2\sin^2(\theta) e^{-\Gamma t'}
\frac{1-\cos\left(\sqrt{4-\Gamma ^2}t'\right)}{4-\Gamma ^2}
\\ \fl
{ \tilde{\rho}}^{10}_{10}(t') =&
\sin ^2(\theta) e^{- \Gamma t' }
\left(
 \frac{\left(2-\Gamma ^2\right) 
       \cos \left(\sqrt{4-\Gamma ^2} t'\right)+2}{4-\Gamma ^2}
 -\frac{\Gamma  \sin \left(\sqrt{4-\Gamma ^2} t' \right)}
       {\sqrt{4-\Gamma ^2}}
\right)
\\
\fl \tilde{\rho}^{00}_{01}(t') =&
i\sin (2 \theta) e^{i\Omega t'-\frac{\Gamma t'}{2}}
\frac{\sin\left(\sqrt{4-\Gamma ^2}\frac{t'}{2}\right)}
 {\sqrt{4-\Gamma ^2}}\\
\fl \tilde{\rho}^{00}_{10}(t')  =&
\frac{\sin (2 \theta)}{2} e^{i  \Omega t' -\frac{ \Gamma t' }{2}}
\left(
 \cos \left(\sqrt{4-\Gamma ^2} \frac{t'}{2}\right)
 -\frac{\Gamma\sin \left(\sqrt{4-\Gamma ^2} \frac{t'}{2}\right)}
  {\sqrt{4-\Gamma ^2}}\right) \\
\fl \tilde{\rho}^{01}_{10}(t')  =&
2i\sin ^2(\theta)e^{-\Gamma t'}
\frac{\sin\left(\sqrt{4-\Gamma^2}\frac{t'}{2}\right)}{\sqrt{4-\Gamma^2}}
\left(
 \frac{\Gamma\sin\left(\sqrt{4-\Gamma^2}\frac{t'}{2}\right)}
      {\sqrt{4-\Gamma^2}}
 -\cos\left(\sqrt{4-\Gamma ^2}\frac{t'}{2}\right)
\right)
\end{eqnarray*}
The expressions of the elements of the density matrix in the critically 
damped case $\Gamma=2$ and in the overdamped case $\Gamma>2$ can be 
obtained from those given in the text for the underdamped case 
$\Gamma<2$. 

\section{Entanglement}
\begin{figure}[htb!]
\begin{center}
$\begin{array}{c@{\hspace{.1in}}c@{\hspace{.1in}}c}
\includegraphics[width=1.8in]{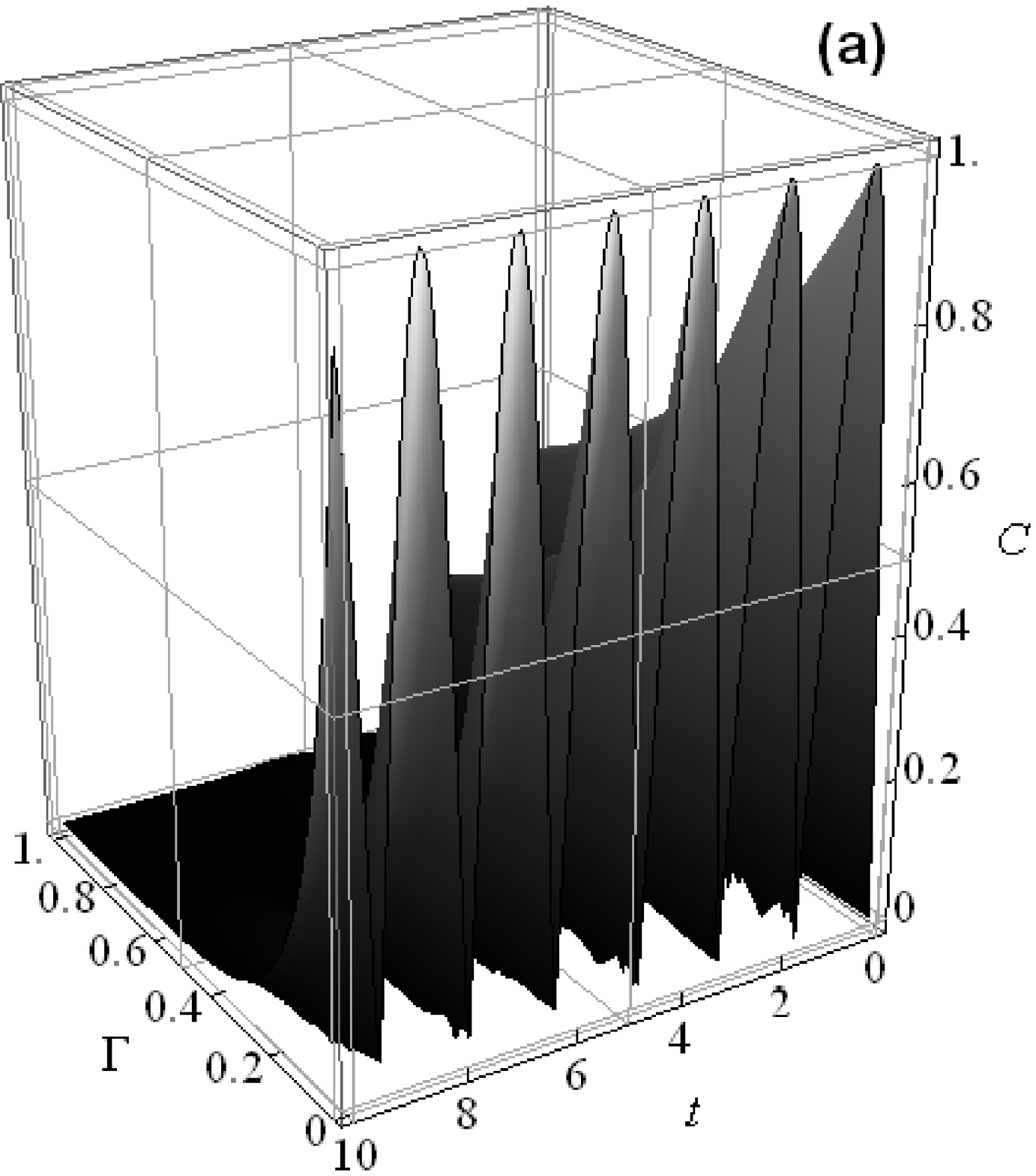}  &
\includegraphics[width=1.9in,height=2in]{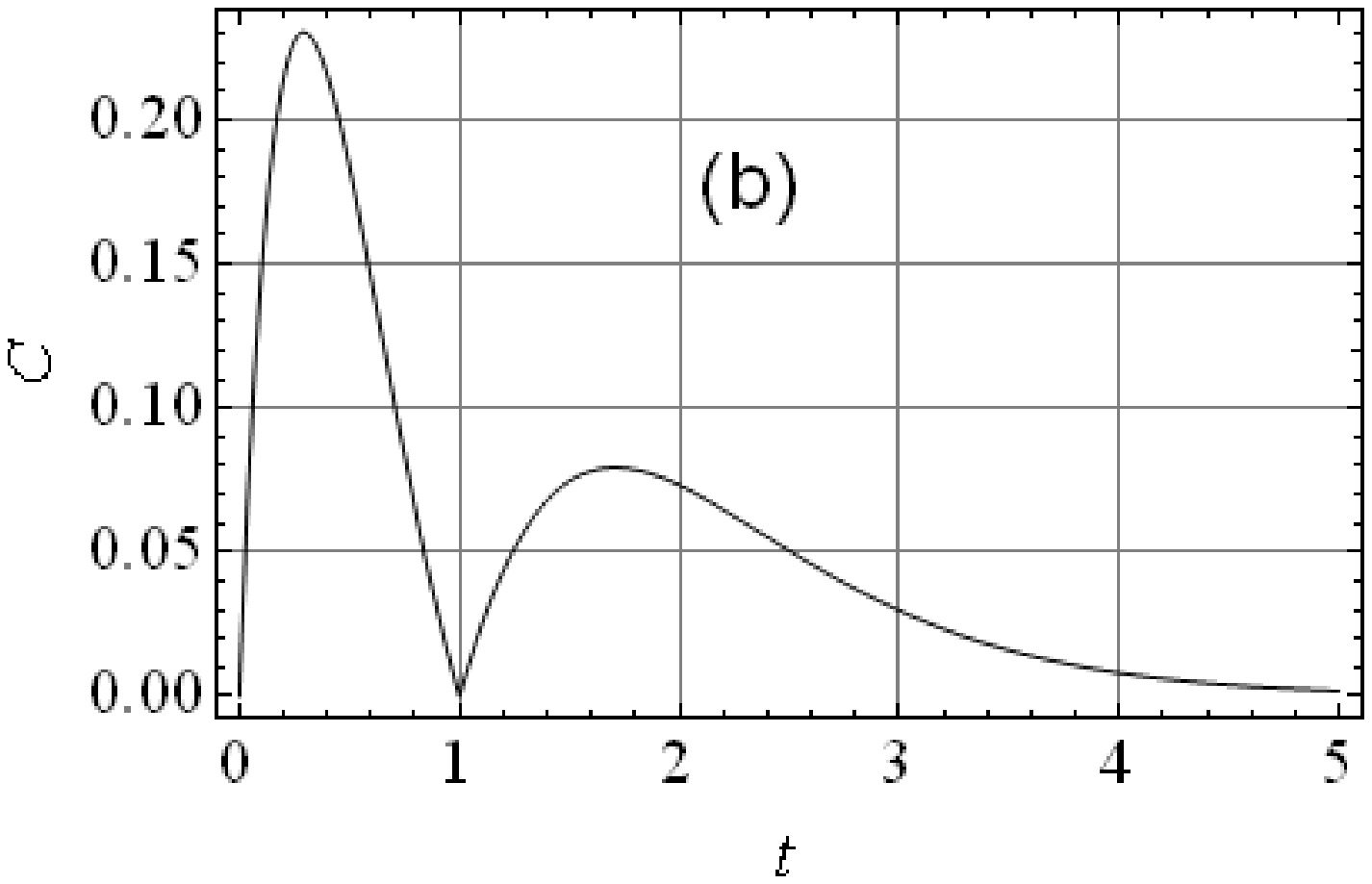} &
\includegraphics[width=1.8in]{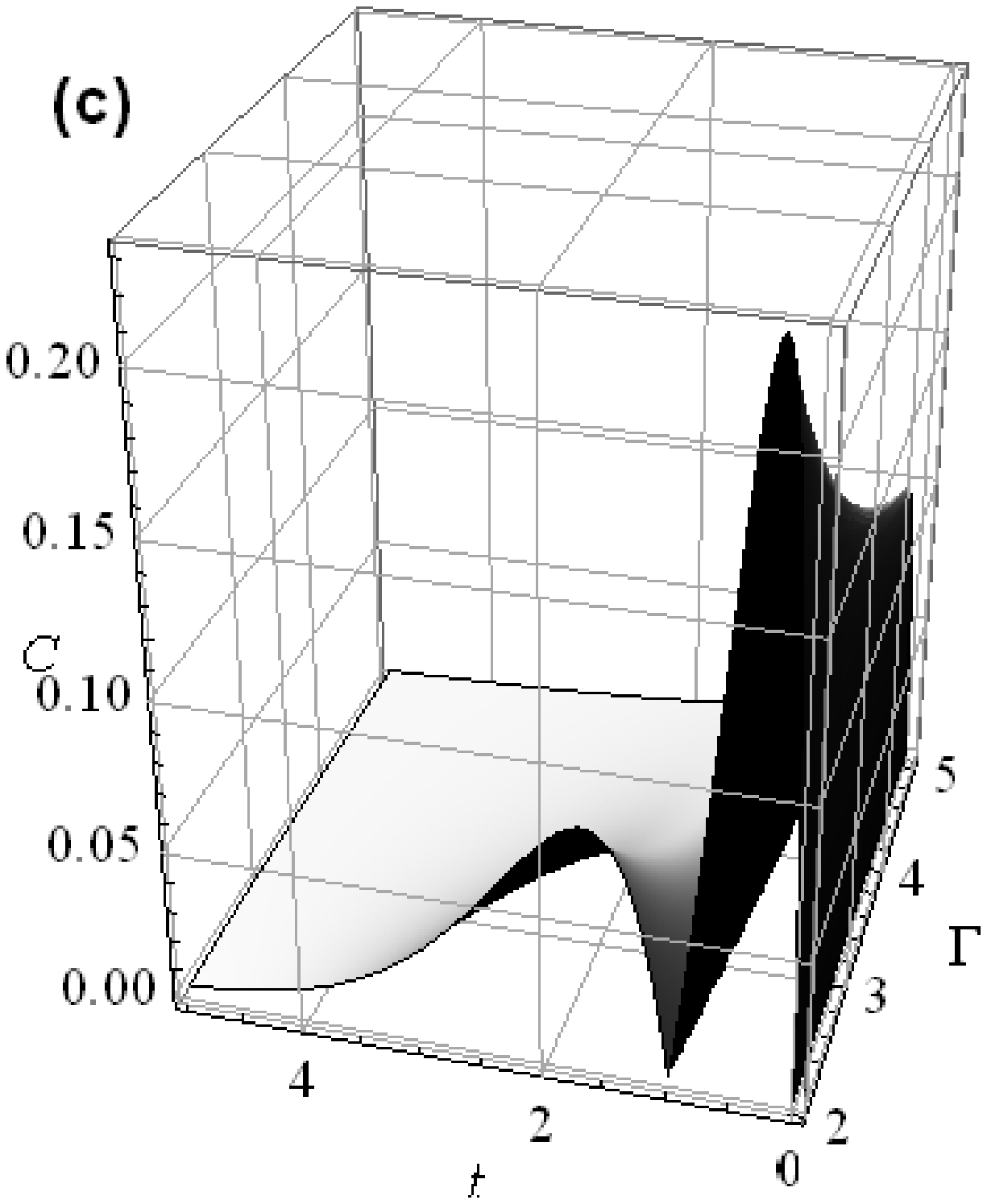}   \\
\end{array}$
\end{center}
\caption{Concurrence as a function of time and the rescaled damping 
constant in the (a) underdamped, (b) critically damped, and 
(c) overdamped case.} 
\label{fig:Concurrence}
\end{figure}

\begin{figure}[h!]
\begin{center}
$\begin{array}{c@{\hspace{.05in}}c@{\hspace{.05in}}c@{\hspace{.05in}}c}
  \includegraphics[width=3.in]{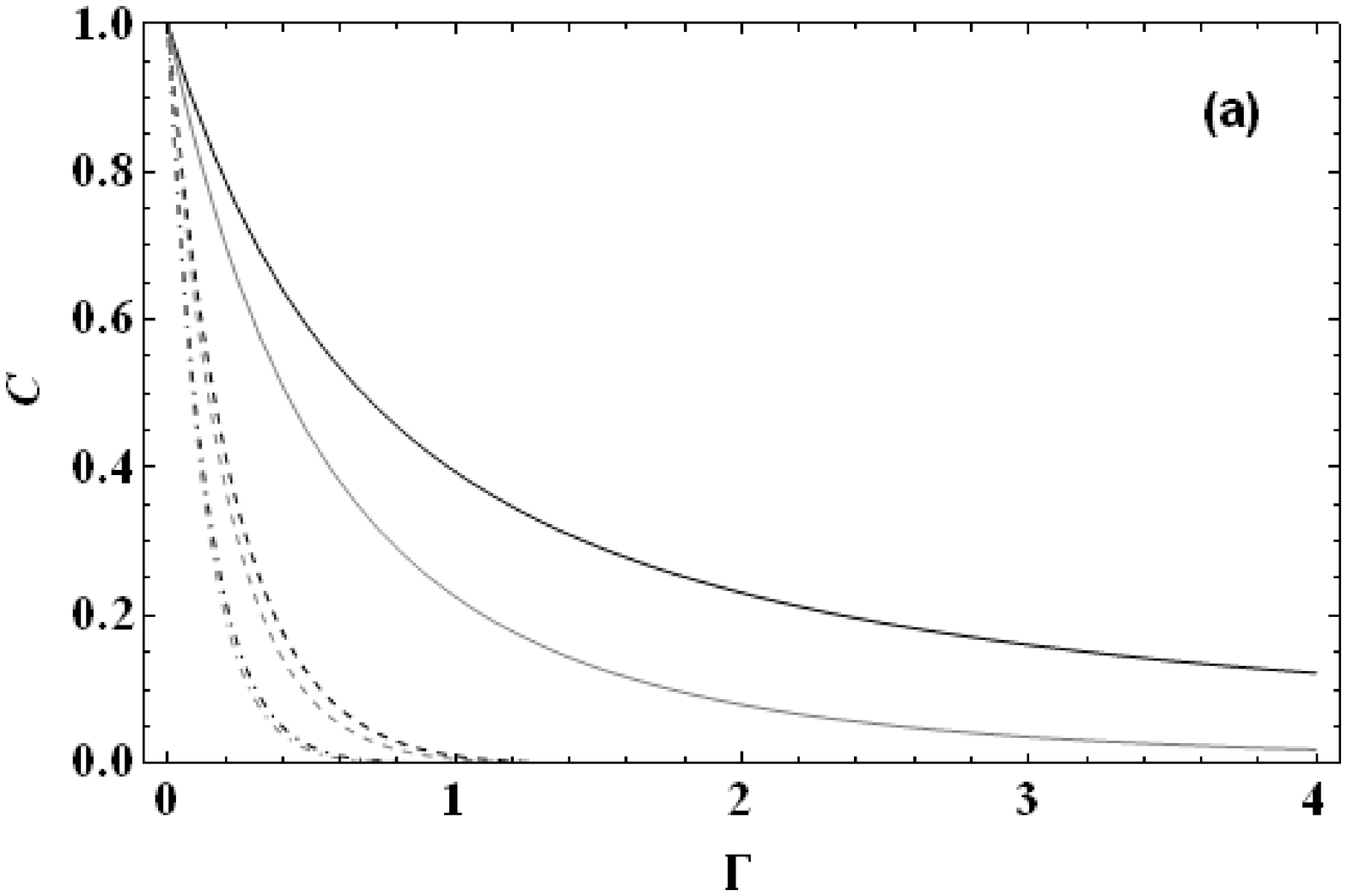} &
  \includegraphics[width=3.in]{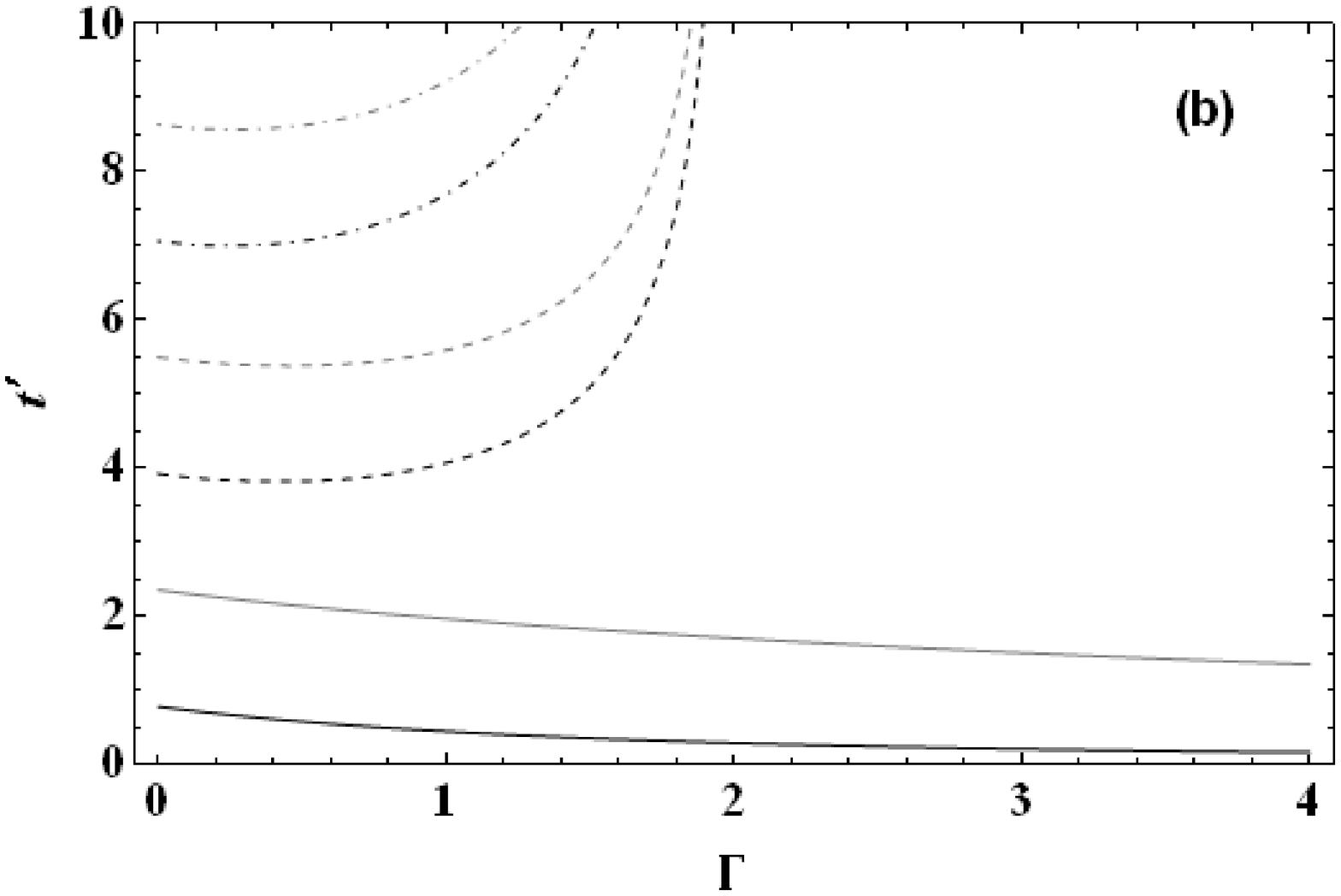}
\end{array}$
\end{center}
\caption{(a) {\bf Concurrence local maxima} at times $\tau_{+(-) n}$ in
black (gray) color for $n=0$ full line, $n=1$ dashed line and $n=2$
dashed-dot line and (b) times of maximum values of concurrence, as a 
function of the rescaled damping constant. } 
\label{fig:MaxConcurrence}
\end{figure}

Although quantities like quantum discord
\cite{Olivier2001a} have been proposed to extract the quantum content
of correlations between two systems, we presently quantify the quantum 
correlations between both oscillators employing a measure of
entanglement. Due to the dynamics of the system, and the initial 
states chosen, the whole system behaves as a couple of qubits and 
therefore its entanglement can be measured by Wootters' concurrence
\cite{Wootters1998a}. 
One of the most important characteristics of the form of the
solution given by (\ref{ansatz}) is that concurrence, as well as
linear entropy, depend \emph{only} on the displaced density
operator $\tilde{\rho}(t')$. In our case the concurrence
reduces to
\begin{eqnarray} \fl
C(t') & =
\left \vert 
 \sqrt{\tilde{\rho}^{01}_{10}(t'){\tilde{\rho}}^{10}_{01}(t')}
 +\sqrt{\tilde{\rho}^{01}_{01}(t') {\tilde{\rho}}^{10}_{10}(t')} 
\right \vert
 -\left\vert 
 \sqrt{{\tilde{\rho}}^{01}_{10}(t')\tilde{\rho}^{10}_{01}(t')}
 -\sqrt{{\tilde{\rho}}^{01}_{01}(t') \tilde{\rho}^{10}_{10}(t')} 
\right\vert
\nonumber \\
\fl
& = 2\sqrt{\tilde{\rho}^{01}_{10}(t')\tilde{\rho}^{10}_{01}(t')} 
= 2 \vert \tilde{\rho}^{10}_{01}(t')\vert,
\end{eqnarray}
where the positivity and hermiticity of the density matrix were used.
The explicit expressions for the concurrence in the underdamped (UD), 
critically  damped (CD) and overdamped (OD) regimes are 
\begin{eqnarray}
\fl
\nonumber C_{UD}(t') &=
4 \sin^2(\theta) e^{-\Gamma t'} 
\frac{\sin \left(\frac{\sqrt{4-\Gamma^2}\ t'}{2}\right)}
     {\sqrt{4-\Gamma^2}}
\left\vert
\frac{\Gamma \sin \left(\frac{\sqrt{4-\Gamma^2}\ t'}{2}\right)}
     {\sqrt{4-\Gamma^2}}
 -\cos \left(\frac{\sqrt{4-\Gamma^2}\ t'}{2}\right)
\right\vert, \\
\fl
C_{CD}(t')& = 2 \sin ^2(\theta ) e^{-2 t'} t'
\big\vert t'-1\big\vert
 , \\
\fl
\nonumber C_{OD}(t') &= 
4 \sin^2(\theta) e^{-\Gamma t'} 
\frac{\sinh \left(\frac{\sqrt{\Gamma^2-4}\ t'}{2}\right)}
     {\sqrt{\Gamma^2-4}}
\left\vert
\frac{\Gamma \sinh \left(\frac{\sqrt{\Gamma^2-4}\ t'}{2}\right)}
     {\sqrt{\Gamma^2-4}}
 -\cosh \left(\frac{\sqrt{\Gamma^2-4}\ t'}{2}\right)
\right\vert .
\end{eqnarray}
All the dependence on the initial state is contained on the squared 
norm of the coefficient of the state $\ket{1}$ of the displaced 
density operator. In all regimes the concurrence vanishes at zero 
time, because the initial state considered is separable. However, 
while in the underdamped case the concurrence  vanishes periodically 
(see equation (\ref{eq:tau2n}) below), in the other two cases it crosses 
zero once ($t>0$) and reaches zero assymptotically as time grows. This 
shows a markedly different qualitative behavior (see figures 
\ref{fig:Concurrence} and \ref{fig:MaxConcurrence}). 

In the underdamped regime the zeroes of the concurrence are found at 
times
\begin{equation}
\label{eq:tau2n}
\tau_{1n}=\frac{2 n \pi }{\sqrt{4-\Gamma ^2}},
\quad \textrm{and} \quad
 \tau_{2n}=
\frac{2 \pi n+2 \arccos \left(\frac{\Gamma}{2}
\right)}{\sqrt{4-\Gamma ^2}},
\end{equation}
where $n$ is a non-negative integer. 
In this contribution, the inverse sine and cosine functions 
are chosen to take values in the interval $[0,\pi/2]$.
The time $\tau_{10}$ corresponds to the initial state.
The sequence of concurrence zeroes is thereby 
$0=\tau_{10}<\tau_{20}<\tau_{11}<\tau_{21}\ldots$
As the critical damping is approached, the time $\tau_{11}$ is pushed 
towards infinity, while $\tau_{20}$ approaches the finite time $2/\Gamma$ 
(see figure \ref{fig:MaxConcurrence}). For the initial states considered 
in this manuscript we do not observe the sudden death of the entanglement 
since the concurrence is zero only for isolated instants of time.

If one writes the concurrence in the underdamped regime in the alternative 
form
\begin{equation}
C_{UD}(t')=
\frac{\sin^2\theta}{2(1-\Gamma^2/4)} e^{-\Gamma t'}
\left|
\frac{\Gamma}{2}
-\sin\left(
     \arcsin\left(\frac{\Gamma}{2}\right)+2\sqrt{1-\frac{\Gamma^2}{4}}\ t'
     \right)
\right|
\end{equation}
it is easy to verify that at the times $\tau_{\pm n}$, given by
\begin{equation}
\fl
\tau_{\pm n} = \frac{1}{\sqrt{4-\Gamma^2}}
\left( 
(2n+1)\pi 
\pm \arccos\left(\frac{\Gamma^2}{4}\right)
-2\arcsin\left(\frac{\Gamma}{2}\right)
\right)
>0,
\quad n=0,1,\ldots
\end{equation}
the concurrence reaches the local maxima
\begin{equation*}
\fl
C_{\pm n} = \sin^2\theta
\left( \sqrt{1+\frac{\Gamma^2}{4}} \pm\frac{\Gamma}{2}\right)
\exp\left( 
 -\frac{\Gamma\left( 
 (2n+1)\pi\pm
  \arccos(\frac{\Gamma^2}{4})-2\arcsin(\frac{\Gamma}{2}) \right)}
{2\sqrt{1-\frac{\Gamma^2}{4}}}
\right) .
\end{equation*}
We observe these maxima to lie on the curves 
$\sin^2\theta K_\pm \exp(-\Gamma t)$, where the constants
$K_\pm =\sqrt{1+\frac{\Gamma^2}{4}} \pm\frac{\Gamma}{2}$ satisfy the 
inequalities 
$\sqrt{2}-1 \leq K_-\leq 1 \leq K_+ \leq \sqrt{2}+1$.
Maxima of concurrence depend on both the initial state and the value 
of the rescaled damping constant, and reach the maximum available value 
of one only in the non-dissipative case for a particular initial state.
In order to have negligible values of concurrence
(except for small time intervals around the zeroes of concurrence)
it is necessary to have times much larger than $1/\gamma$.
From the point of view of classical-like behavior, the most favourable
scenario corresponds to zero or almost zero concurrence,
which are obtained for short time intervals around 
$\tau_{1 n},\ \tau_{2 n}$ and for large values of time.

In the overdamped regime, the concurrence presents two maxima, $\tau_-$ 
and $\tau_+>\tau_-$
\begin{equation}
\tau_\pm = 
\frac{2 \textrm{arccosh}(\Gamma/2)\pm \textrm{arccosh}(\Gamma^2/4)}
{2 \sqrt{1-\Gamma^2/4}},
\end{equation}
both of which go to zero as the rescaled dissipation rate
grows, $\tau_+\rightarrow 4 \ln(\Gamma)/\Gamma$ and 
$\tau_-\rightarrow \ln(2)/\Gamma$
(see figure \ref{fig:MaxConcurrence}).
The function arccosh$(x)$ is chosen as to return nonnegative values for 
$x\geq 1$. Since the  global maximum of concurrence, which corresponds 
to the later time, scales like $1/(2\Gamma)$ for large values of $\Gamma$, 
in the highly overdamped regime quantum correlations are not developed 
at any time.

The behavior of concurrence in the different regimes is shown in figure 
\ref{fig:Concurrence}. It is apparent that small values of concurrence 
are obtained for very small times and for large times in the underdamped 
case and for all times for a highly overdamped oscillator. In figure 
\ref{fig:MaxConcurrence} we depict the times at which concurrence attains 
a maximum, and the maximum values of concurrence, as a function of the 
rescaled damping constant. 
One can see how the first two times of maximum concurrence go to zero,
while the other times diverge, as the critically damped regime is reached. 
The first two maxima of concurrence vanish more slowly than the rest of 
maxima, which hit zero at $\Gamma=2$.

\section{Entropy}
The entropy is analized employing the linear entropy of the first oscillator,
the system of interest. As remarked before, the first oscillator behaves like 
a two-level system, where the maximum value of the linear entropy, 0.5, is obtained 
when the population of each of the two states is one half. 
The type of ``classical'' behavior which allows the interaction with Ramsey 
zones to be modelled like a classical driving force occurs when the linear 
entropy is very small, and hence the state of the first oscillator is (almost) 
pure and uncorrelated with the state of the second oscillator.
The  linear entropy for the first oscillator can be computed as
\begin{equation}
\label{eq:delta1}
\delta_{1}(t')
= 1- \tr_1 \left( \tr_2 \rho \tr_2 \rho  \right)
= 1- \tr_1 \left( \tr_2 \tilde{\rho} \tr_2 \tilde{\rho}  \right)
= 2 \det \left(\tr_2 \tilde{\rho}\right),
\end{equation}
where the last equality holds for two-level systems. In equation (\ref{eq:delta1}) 
the density operator of the first oscillator is assumed to be represented by a
2$\times$2 matrix. Employing the expressions we have found for the elements of 
$\tilde{\rho}$ we obtain
\begin{equation}
\label{eq:delta}
\delta_{1}(t')
= 2 \sin^4\theta\ x(t') 
\left(1 - x(t')
 \right),
\end{equation}
where $x$, in the underdamped regime, is given by
\begin{equation}
x_{UD}(t)=
\frac{ e^{-\Gamma t}\sin^2\left(\sqrt{1-\Gamma^2/4}\ t-\arccos(\Gamma/2)\right)}
     {1-\Gamma^2/4}.
\end{equation}
Surprisingly, as in the case of concurrence, the influence of the initial state 
factors out in the expression of the linear entropy of the first oscillator, which
turns out to be proportional to the square of the population of 
$\ket{1}$ in the initial displaced operator. 
As it is well known, in the limit of zero 
dissipation, the linear entropy of the reduced density matrix is equal to one fourth
of the square of the concurrence.
At times $\tau_{2n}$ (see eq.(\ref{eq:tau2n})), when both concurrence and 
linear entropy vanish, the total state of the system is separable,
$\rho(g\tau_{2n}) = \ket{\beta(g\tau_{2n})}\bra{\beta(g\tau_{2n})}
\otimes \rho_2(g\tau_{2n})$; that is, from the point of view of the first
oscillator the evolution is unitary like.
Since the linear entropy begins at zero, because the initial state is pure and 
separable, there is a maximum in the interval $(0,\tau_{20})$, which turns out to
give a linear entropy of exactly 0.5 (we treat the case $\sin\theta=1$, because ---due 
to the scaling property discussed before--- a simple multiplication by $\sin^4\theta$
gives the result for other cases).
Indeed, as the function $x(t)$ changes continuously from $x(t=0)=1$ to $x(t=\tau_{20})=0$, 
it crosses 0.5 at some time $\tau_{30}$ in between, giving the maximum value possible of 
the linear entropy. Although the exact value of $\tau_{30}$ can be obtained only 
numerically, good analytical approximations can be readily obtained. For example, 
$\tau_{30}\approx\pi/(4+4g+2g^2)$, gives an error smaller than 0.5\%.

For small values of the rescaled damping constant, $\Gamma\lessapprox 0.237$, 
there are several solutions to the equation $x(t)=0.5$ in the interval $(0,\log(2)/\Gamma)$,
which give absolute maxima of the linear entropy, while the times
\begin{equation}
\label{eq:tau4n}
\tau_{4n} = \frac{2\arccos(\Gamma/2)+n\pi}{\sqrt{1-\Gamma^2/4}},
\quad
n=0,1,2,\cdots
\end{equation}
correspond to local minima. In the interval $(\log(2)/\Gamma,\infty)$ the times
$ \tau_{4n}$ give local maxima. All of the local maxima and minima given by 
eq. (\ref{eq:tau4n}) belong to the curve $2 e^{-\Gamma t}(1-e^{-\Gamma t})$. 
The large time behavior of the local maxima of linear entropy and concurrence
is, thereby, of the same form constant$\times \exp(-\Gamma t')$.  
For values of $\Gamma>0.237$ all times $ \tau_{4n}$ give local maxima.
The maxima of concurrence and linear entropy coincide only in the weakly 
damped case, because concurrence and linear entropy are not independent 
for pure bipartite states.

At times $\tau_{1n}$ (see eq.(\ref{eq:tau2n})), where the total state 
$\rho(g\tau_{1n}) = \rho_1(g\tau_{1n}) \otimes 
\ket{\alpha(g\tau_{1n})}\bra{\alpha(g\tau_{1n})}$, is separable, the reduced state
of the first oscillator is mixed. The linear entropy is small for short 
$(\tau_{1n}\ll \log(2)/\Gamma)$ and large $(\tau_{1n}\gg \log(2)/\Gamma)$ times.

\begin{figure}[h!]
\begin{center}
$\begin{array}{c@{\hspace{.1in}}c@{\hspace{.1in}}c}
\includegraphics[width=1.7in]{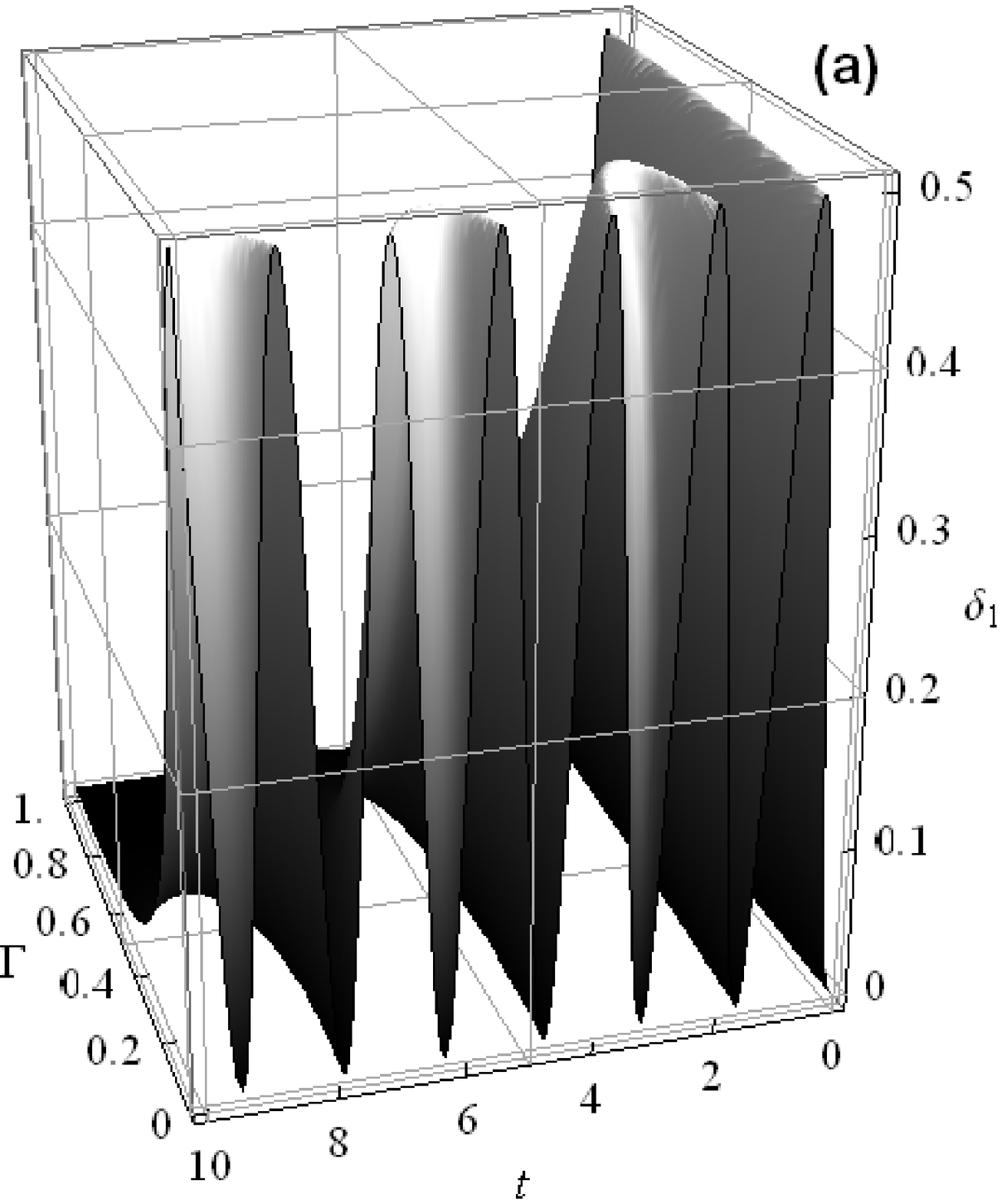}  &
\includegraphics[width=1.75in,height=2in]{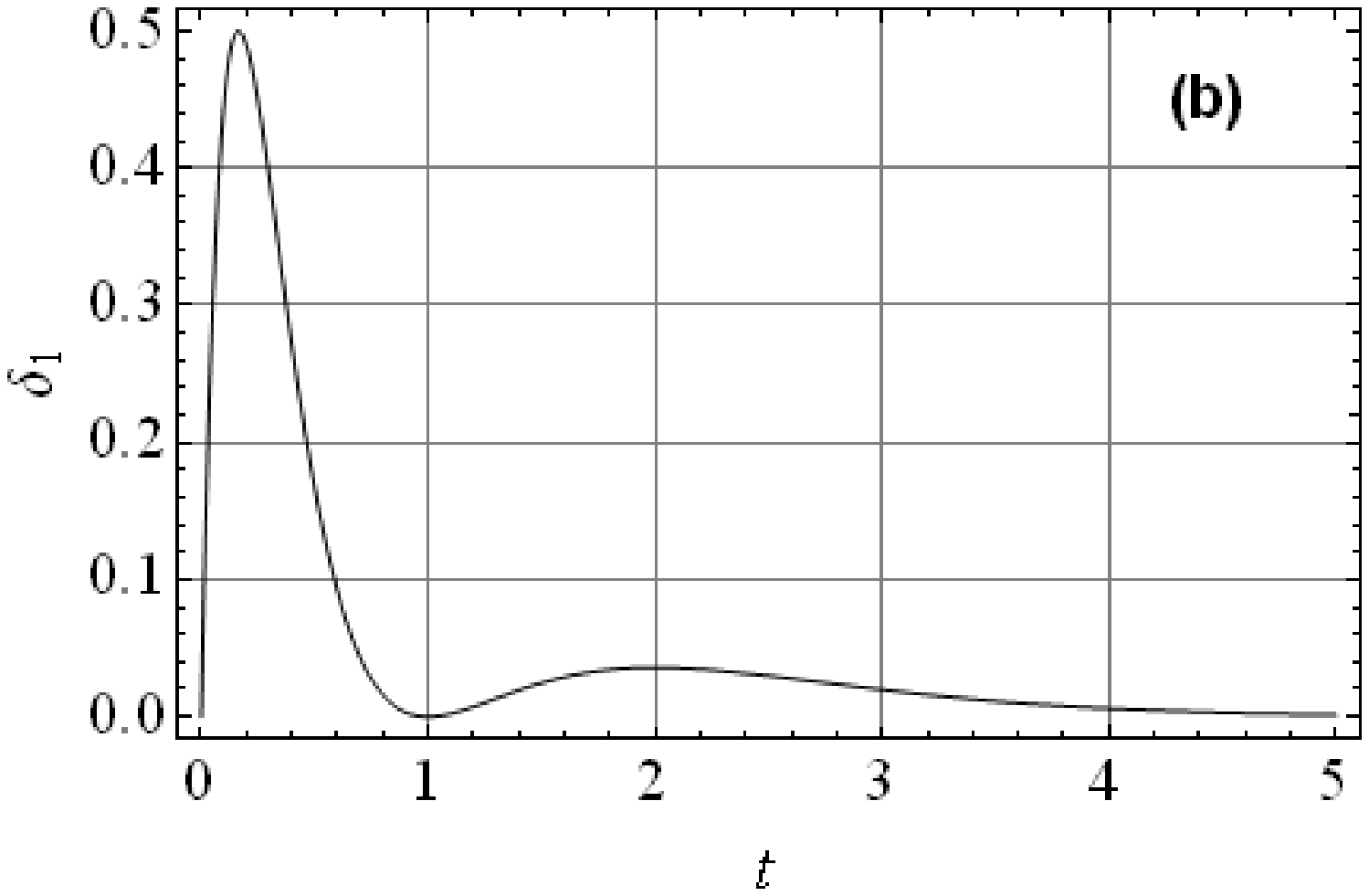} &
\includegraphics[width=1.7in]{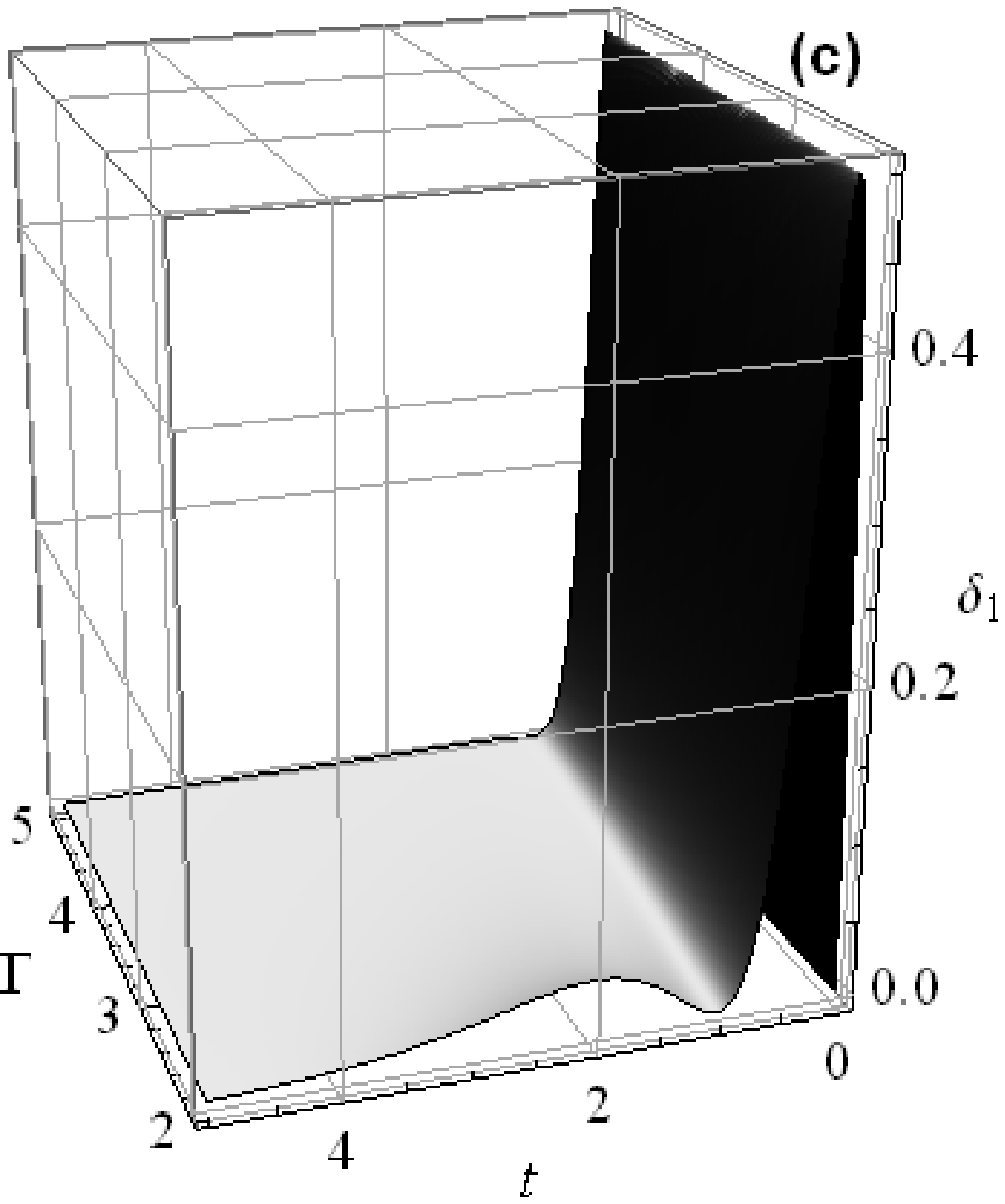}   \\
\end{array}$
\end{center}
\caption{Linear entropy of the first oscillator as a function of time 
and the rescaled damping constant in the 
(a) underdamped, (b) critically damped, and (c) overdamped case.} 
\label{fig:LinearEntropy}
\end{figure}

In the overdamped regime 
the function $x(t)$, which appers on the expression for linear entropy 
(\ref{eq:delta}) and is given by
\begin{equation}
x_{OD}(t')=
\frac{e^{-\Gamma t'} \sinh^2(\sqrt{\Gamma^2/4-1}\ t'-\textrm{arccosh}(\Gamma/2))}
{\Gamma^2/4-1},
\end{equation} 
begins at one for $t'=0$, and goes down to zero for large values of time.
The time at which it crosses one half can be calculated to be 
$\tau_{0.5}\approx 0.16557 < 1/6$ for $\Gamma=2$ and for large values
of $\Gamma$ it goes as $\tau_{0.5}\approx \ln 2/(2\Gamma)$. It is easy
to find interpolating functions with small error for the time of crossing,
\begin{equation*}
\tilde{\rho}_{10}^{10}
\left(t'=\frac{1}{
  6 +\frac{4}{\ln 2}
  \sinh\left(
    \textrm{arccosh}(\frac{\Gamma}{2})
    \tanh\left(\frac{\textrm{arccosh}(\frac{\Gamma}{2})}{1.6}\right)
  \right)}
\right)=
0.5(1+\Delta)
\end{equation*} 
where $|\Delta|<2.5$\%. It is interesting to notice that for large
values of the damping this time ($\ln(2)/(2\Gamma)$) is half the 
time needed to obtain the 
maximum value of concurrence, and that, at the later time, the linear entropy 
is 3/4 of the maximum value of entropy, a relatively large value. The state
of the first oscillator always becomes maximally mixed before becoming
pure again, no matter how large the value of the damping. We show the behavior of
linear entropy in figure \ref{fig:LinearEntropy}. In the underdamped regime 
there are infinite maxima and minima, while for critical damping and
for the overdamped regime there are only two maxima. The first maximum
always corresponds to a linear entropy of one half.

\section{Conclusions}
In the present contribution we have shown that the classical quantum
border in this model depends mainly on the initial state and on damping 
constant to interaction coupling ratio, and that quantum effects, 
characteristic of the underdamped regime, can be seen in the other 
regimes for small times. In order to make connection with Ramsey zones
we remember in that physical system $\omega \approx 10^{10}$ Hz, 
$Q \approx 10^4$, $g \approx 10^4$ Hz and  $T_R \approx 10^{-5}$ s, which
was chosen as to produce $\pi/2$ pulse, that is a pulse that 
can rotate the state of the two-level system, as represented in a Bloch
sphere, by an angle $\pi/2$.
These numbers place the system into the highly overdamped (regime because 
$\Gamma = \omega/(Q g) \approx 10^2 \gg 2$) and give a rescaled evolution
time of the order of $g T \approx 10^{-1}$.
Here we use the same values of $\omega, \gamma$ and $g$, and an evolution 
time of order 1/g. Indeed, the hamiltonian 
$ \hbar\omega{\hat b}^{\dagger}{\hat b}
+  \hbar g \left(\Theta(t)-\Theta(t+T)\right)
(\alpha_0 e^{-i\omega t}{\hat b}^{\dagger}
+\alpha_0^* e^{i\omega t}{\hat b})$,
with $\|\alpha_0\|\approx 1$ --- which would model the interaction of the 
first oscillator with a classical driving field of an average number of 
excitations of the order of one ---
has a characteristic time $1/g$, corresponding to $T'\approx 1$. 

The dynamical behavior of the linear entropy obtained here, is quite 
different from that of ref. \cite{Kim1999a}: there the linear entropy 
was never large for the relevant time interval, here it grows to the 
maximum possible for a two-level system, and then goes to zero very 
quickly. Therefore, in this model dissipation produces relaxation also, 
and a description obviating the second oscillator still needs a dissipation 
process. Although, at the evolution time T, both models predict a small 
atomic entropy, in Ramsey zones it decreases as 
$\delta_1({T_R}'\approx 0.1)\approx 4/\Gamma$, while in the present model 
it  goes to zero as $\delta_1(T'\approx 1)\propto 1/\Gamma^4$.
Qualitative and quantitative differences notwithstanding, at the evolution 
time the linear entropy is very small, in both cases, due to the smallness 
of the ratio $g/\gamma$. 
As remarked before the quality factor of the damped oscillator does
not appear directly in either case; it can be perfectly possible to have a 
very weakly damped oscillator and a highly overdamped interaction. However, as the first 
oscillator quality factor is improved, the damping constant will eventually be
comparable with the interaction constant, and there will be considerable entanglement
between both oscillators. For the same physical system if the damping rate
can be changed then classical or quantum behavior can be obtained.

\section*{Acknowledgements}
This work was partially funded by DIB-UNAL and Facultad de Ciencias, Universidad Nacional
(Colombia).

\section*{References}
%

\end{document}